\documentclass[11pt,a4paper]{article}

\pdfoutput=1

\usepackage{comment}
\usepackage{jcappub}

\usepackage{amsmath}

\usepackage{physics}
\usepackage{subfig}
\usepackage{float}
\usepackage{xcolor}

\newcommand{\arcsinh}{\text{arcsinh}}
\newcommand{\be}{\begin{equation}}
\newcommand{\ee}{\end{equation}}
\newcommand{\bea}{\begin{eqnarray}}

\newcommand{\eea}{\end{eqnarray}}
\newcommand{\nn}{\nonumber}

\newcommand{\MP}{M_\text{P}}

\newcommand{\phimin}{\phi_\text{min}}
\newcommand{\phimax}{\phi_\text{max}}

\title{\centering $\tilde\xi$-attractors in metric-affine gravity}

\author[a]{A. Racioppi}

\affiliation[a]{National Institute of Chemical Physics and Biophysics, R\"avala 10, 10143 Tallinn, Estonia}

\emailAdd{antonio.racioppi@kbfi.ee}

\abstract{We propose a new class of inflationary attractors in metric-affine gravity. Such class features a non-minimal coupling $\tilde\xi \, \Omega(\phi)$ with the Holst invariant $\tilde {\cal R}$ and an inflaton potential proportional to $\Omega(\phi)^2$. The attractor behaviour of the class takes place with two combined strong coupling limits. The first limit is realized at large $\tilde\xi$, which makes the theory equivalent to a $\tilde {\cal R}^2$ model. Then, the second limit considers a very small Barbero-Immirzi parameter which leads the inflationary predictions of the $\tilde {\cal R}^2$ model towards the ones of Starobinsky inflation. Because of the analogy  with the renown $\xi$-attractors, we label this new class as $\tilde\xi$-attractors.}


\begin{document}
\maketitle

\section{Introduction}\label{intro}
Cosmic inflation, i.e. an accelerated expansion during the very early Universe, is the current paradigm for explaining the flatness and homogeneity of the Universe at large scales~\cite{Starobinsky:1980te,Guth:1980zm,Linde:1981mu,Albrecht:1982wi}. Moreover, it also provides an origin for  the small inhomogeneities observed in the Cosmic Microwave Background radiation. In its minimal version, inflation is usually formulated by adding to the Einstein-Hilbert action one scalar field, the inflaton, whose energy density induces a near-exponential expansion. 

The latest combination of Planck, BICEP/Keck and BAO data~\cite{BICEP:2021xfz} has sensibly reduced the allowed parameters space, strongly favouring nearly-flat concave inflaton potentials and already ruling out many proposed models. Nevertheless, the most popular inflationary realizations, the Starobinsky model~\cite{Starobinsky:1980te} and Higgs-inflation~\cite{Bezrukov:2007ep}, still sit in the allowed region. Both models can be described by a scalar field non-minimally coupled to gravity  (e.g.~\cite{Jarv:2016sow} and references therein).

However, when theories are non-minimally  coupled to gravity there is more than one \emph{choice} of the dynamical degrees of freedom. 
In  the  more popular metric gravity, the  metric tensor is the only dynamical degree of freedom, while the connection is fixed to be the Levi-Civita one.  On the other hand,  in metric-affine gravity (MAG), both the metric and the connection are dynamical variables and their corresponding equations of motion will dictate the eventual relation between them. When the gravity action contains only the term linear in the curvature scalar and no fermions, the two approaches lead to equivalent theories (e.g.~\cite{BeltranJimenez:2019esp,Rigouzzo:2023sbb} and refs. therein), otherwise the theories are completely different~\cite{BeltranJimenez:2019esp,Rigouzzo:2023sbb,Koivisto:2005yc,Bauer:2008zj} and lead to different phenomenological predictions, as recently investigated in e.g.~\cite{Tamanini:2010uq,Bauer:2010jg,Rasanen:2017ivk,Tenkanen:2017jih,Racioppi:2017spw,Markkanen:2017tun,Jarv:2017azx,Racioppi:2018zoy,Kannike:2018zwn,Enckell:2018kkc,Enckell:2018hmo,Rasanen:2018ihz,Bostan:2019uvv,Bostan:2019wsd,Carrilho:2018ffi,Almeida:2018oid,Takahashi:2018brt,Tenkanen:2019jiq,Tenkanen:2019xzn,Tenkanen:2019wsd,Kozak:2018vlp,Antoniadis:2018yfq,Antoniadis:2018ywb,Gialamas:2019nly,Racioppi:2019jsp,Rubio:2019ypq,Lloyd-Stubbs:2020pvx,Das:2020kff,McDonald:2020lpz,Shaposhnikov:2020fdv,Enckell:2020lvn,Jarv:2020qqm,Gialamas:2020snr,Karam:2020rpa,Gialamas:2020vto,Karam:2021wzz,Karam:2021sno,Gialamas:2021enw,Annala:2021zdt,Racioppi:2021ynx,Cheong:2021kyc,Mikura:2021clt,Ito:2021ssc,Racioppi:2021jai,Lillepalu:2022knx,Gialamas:2023flv,Piani:2023aof,Barker:2024ydb,Dioguardi:2021fmr,Racioppi:2022qxq,Dioguardi:2022oqu,Dioguardi:2023jwa,Kannike:2023kzt,TerenteDiaz:2023kgc,Dimopoulos:2022rdp,SanchezLopez:2023ixx,Marzo:2024pyn}. Moreover, MAG admits two, rather than just one, two-derivative curvature invariants: the usual Ricci-like scalar and the Holst invariant~\cite{Hojman:1980kv,Nelson:1980ph,Holst:1995pc}, which can be used to construct new models~(e.g.~\cite{Hecht:1996np,BeltranJimenez:2019hrm,Langvik:2020nrs,Shaposhnikov:2020gts,Pradisi:2022nmh,Salvio:2022suk,Piani:2022gon,DiMarco:2023ncs,Gialamas:2022xtt,Gialamas:2024jeb,Gialamas:2024iyu,Racioppi:2024zva}).

The purpose of this article is to study a new model in MAG, where the Jordan frame inflaton scalar potential is proportional to the square of the non-minimal coupling function involving the Holst invariant and/or the Ricci-like curvature scalar. As we will see later, this kind of setup will induce a new class of inflationary attractors.

The discussion is organized as follows. In Section~\ref{sec:model} we introduce the action for our inflationary model in metric-affine gravity. For the sake of minimality, we consider only constructions where the physical degrees of freedom are only the graviton and the inflaton and without terms involving more than two derivatives. In Section~\ref{sec:U} we present an analytical study of the inflaton potential and describe how the attractor configuration is reached. Then, in Section~\ref{sec:results} we present a detailed numerical study with the corresponding inflationary predictions. Finally, in Section~\ref{sec:conclusions} we summarize our conclusions. In addition, in Appendix~\ref{appendix:attractors} we provide more details about the attractor configurations, while in Appendix~\ref{appendix}, we show the full analytical equations of the slow-roll parameters and the inflationary observables.

\section{Model} \label{sec:model}
We start with the Jordan frame action for a real scalar $\phi$ non-minimally coupled to gravity
\be 
S_{\rm J}= \int d^4x\sqrt{-g}\left[\frac{M_P^2}{2} \left( f(\phi){\cal R}+\tilde f(\phi)\tilde{\cal R} \right)  -\frac{\partial_\mu \phi \, \partial^\mu \phi}{2} - V(\phi) \right], 
\label{eq:Sstart} 
\ee
where $M_P$ is the reduced Planck mass, $V(\phi)$ the inflaton potential,  $f(\phi)$ and $\tilde f(\phi)$ are non-minimal coupling functions, ${\cal R}$ and $\tilde{\cal R}$  respectively, a scalar and pseudoscalar  contraction of the curvature (the latter also known as the Holst invariant~\cite{Hojman:1980kv,Nelson:1980ph,Holst:1995pc}),
\be 
{\cal R} \equiv {\cal F}_{\mu\nu}^{~~~\mu\nu}, \qquad  \tilde{\cal R} \equiv \frac1{\sqrt{-g}}\epsilon^{\mu\nu\rho\sigma}{\cal F}_{\mu\nu\rho\sigma},\label{eq:RRpdef}
\ee
where $\epsilon^{\mu\nu\rho\sigma}$ is the totally antisymmetric Levi-Civita symbol with $\epsilon^{0123}=1$. ${\cal F}_{\mu\nu~~\sigma}^{~~~\rho}$ is the curvature associated with the connection ${\cal A}_{\mu~\sigma}^{~\,\rho}$,
\be 
{\cal F}_{\mu\nu~~\sigma}^{~~~\rho} \equiv \partial_\mu{\cal A}_{\nu~\sigma}^{~\,\rho}-\partial_\nu{\cal A}_{\mu~\sigma}^{~\,\rho}+{\cal A}_{\mu~\lambda}^{~\,\rho}{\cal A}_{\nu~\sigma}^{~\,\lambda}-{\cal A}_{\nu~\lambda}^{~\,\rho}{\cal A}_{\mu~\sigma}^{~\,\lambda} \ .
\ee
As mentioned before, we do not consider any other term in action~\eqref{eq:Sstart} in order to keep the model as minimal as possible, with only the massless graviton and the inflaton as physical degrees of freedom  (no $\tilde{\cal R}^2$ term (e.g. \cite{Salvio:2022suk} and refs. therein)) and without terms that feature more than two derivatives (no ${\cal R}^2$ like terms (e.g. \cite{Annala:2021zdt} and refs. therein)).

We remind that in MAG, the connection  ${\cal A}_{\mu~\sigma}^{~\,\rho}$ is not assumed to be the Levi-Civita one, but it is computed from the corresponding equation of motion. We also remind that, if ${\cal A}_{\mu~\sigma}^{~\,\rho}$ is the Levi-Civita connection,  $\tilde{\cal R}$ vanishes\footnote{The careful reader might notice that $\tilde{\cal R}$ actually vanishes for any  ${\cal A}_{\mu~\sigma}^{~\,\rho}$ that satisfies  ${\cal A}_{\mu~\sigma}^{~\,\rho} =  {\cal A}_{\sigma~\mu}^{~\,\rho}$.} and ${\cal R}$ equals the Ricci scalar $R$.
After some manipulations (e.g.~\cite{Racioppi:2024zva} and refs. therein), the action~\eqref{eq:Sstart} can be written in the Einstein frame as
\be 
S_{\rm E} =\int d^4x\sqrt{-g}\left[\frac{\MP^2}{2} R -\frac{1}{2}\partial_\mu \chi \, \partial^\mu \chi - U(\chi) \right],
\label{eq:SE}
\ee 
where the Einstein frame scalar potential is
\be
 U(\chi)=\frac{V(\phi(\chi))}{f^2(\phi(\chi))} \label{eq:U} \, ,
\ee
\be
  \left( \frac{d\chi}{d\phi} \right)^2= k(\phi) \, , \qquad k(\phi) = \frac1f+\frac{6 M_P^2 \left[f(\phi)' {\tilde f(\phi)}-f(\phi){\tilde f(\phi)}'\right]^2}{f(\phi)^2 \left[ f(\phi)^2+4{\tilde f(\phi)}^2 \right]},
    \label{eq:k(phi)}
\ee
where $'$ represents a derivative with respect to argument of the function. Since $f(\phi)$ must be positive in order to avoid repulsive gravity, $k(\phi)$ is always positive and $\phi$ never a ghost.

In the present article we are interested in the $\xi$-attractors-inspired~\cite{Kallosh:2013tua,Galante:2014ifa} configuration
\be 
V(\phi) = \Lambda^4 \, \Omega(\phi)^2, \qquad
f(\phi) = 1 + \xi \, \Omega(\phi), \qquad 
{\tilde f}(\phi) = {\tilde f}_0^2 + \tilde \xi \, \Omega(\phi) \, , \label{eq:attractorschoice}
\ee 
where $\Omega$ is a positive continuous and differentiable function of $\phi$.
The quantity $1/(4{\tilde f}_0^2)$ is known as the  Barbero-Immirzi parameter
~\cite{Immirzi:1996di,Immirzi:1996dr}. 
Similar setups have been already studied in the literature.
For instance, the configuration with $\Omega \propto \phi^2$ has been studied in~\cite{Langvik:2020nrs,Shaposhnikov:2020gts}. According to our knowledge, for what concerns the last configuration, the setup with a generic $\Omega$ has not been studied yet, which is instead the purpose of our work.  
Moreover, the case  with $f \neq 1$ but $\tilde{f}_0=\tilde\xi=0$ has been already studied respectively in the context of metric gravity (i.e. the aforementioned $\xi$-attractors \cite{Kallosh:2013tua,Galante:2014ifa}) and MAG (i.e. Palatini gravity, since $\tilde f =0$) in \cite{Bauer:2008zj,Jarv:2017azx}. In metric gravity the attractor behaviour of \cite{Kallosh:2013tua,Galante:2014ifa}  was the combined results of two universal asymptotic strong coupling ($\xi \gg 1$) limits: the Einstein frame potential (which has the same form as eq. \eqref{eq:U}) that converges to a constant and the canonical normalization of the inflaton (cf. eq. \eqref{eq:k(phi)}), with an additional $\frac{f'}{f}$ contribution that becomes dominant and converges to a universal $\ln(f)$ behaviour when integrated. The combination of the two gives Starobinsky inflation as the attractor solution. On the other hand, in the Palatini realization \cite{Bauer:2008zj,Jarv:2017azx}, while the Einstein frame potential limit remains universal, the field redefinition is not (the $\frac{f'}{f}$ term is absent), implying a loss of attractor behaviour\footnote{More details are available in \cite{Jarv:2020qqm} and appendix \ref{appendix:attractors}.}. Therefore, in this article, since we look  for attractors solution, 
we study in details the setup where $\Omega$ is a generic function, but only with  $\xi=0$ (i.e. only a non-minimal coupling between $\phi$ and the Holst invariant), leaving the most general study for a future work\footnote{We will provide an additional small comment about the impact of $\xi > 0$ at the end of subsection \ref{subsec:validity}.}.

\section{General features of the scalar potential} \label{sec:U}
In the case of $f(\phi)=1$ i.e. $\xi=0$ (see eq.~\eqref{eq:attractorschoice}), the scalar potential in eq.~\eqref{eq:U} just becomes $U(\chi)= V(\phi(\chi))$. Therefore, the eventual change of shape in the potential is all due to the field redefinition in eq.~\eqref{eq:k(phi)}. When $\xi=0$, the non-minimal kinetic function simply becomes
\be
   k(\phi) = 1+\frac{6 M_P^2 \left[{\tilde f(\phi)}'\right]^2}{ \left[ 1+4{\tilde f(\phi)}^2 \right]}
   = 1+\frac{6 M_P^2 \, \tilde{\xi }^2 \left[\Omega(\phi)'\right]^2}{ 1+4 \left(\tilde{f_0}^2+\tilde{\xi } \Omega(\phi) \right)^2} \, .
    \label{eq:k(phi):omega}
\ee
%
It is well known that when $k(\phi)$ presents a pole (or just a pronounced peak), $U(\chi)$ exhibits a flat region that might be suitable for inflation. Since the denominator in eq.~\eqref{eq:k(phi):omega} is strictly positive, the chance of a pole is excluded and we are left only with the possibility of an eventual local maximum with $k(\phi) \gg 1$. This should naively happen when $|\tilde\xi| \gg 1$ (see eq.~\eqref{eq:k(phi):omega}). In such a case, then  we can easily approximate the behaviour of $k(\phi)$ nearby the maximum by neglecting the ``$1+$" term before the fraction in~\eqref{eq:k(phi):omega}, obtaining
\be
   k(\phi) \simeq \frac{6 M_P^2 \, \tilde{\xi }^2 \left[\Omega(\phi)'\right]^2}{ 1+4 \left(\tilde{f_0}^2+\tilde{\xi } \Omega(\phi) \right)^2} \, .
    \label{eq:k(phi):omega:app}
\ee
Using such an expression, we can provide an approximated solution for~\eqref{eq:k(phi)}
\be
 \chi \simeq -\sqrt{\frac{3}{2}} M_P \left\{ \arcsinh\left[2 \left(\tilde{f_0}{}^2+\tilde{\xi } \Omega(\phi)\right)\right]- \arcsinh\left(2 \tilde{f_0}{}^2\right) \right\} \, ,
   \label{eq:chi:omega:app}
\ee
which can be inverted in function of $\Omega(\phi)$, allowing us to write explicitly the Einstein frame potential as
\be
 U(\chi)_{\tilde {\cal R}^2} \simeq \frac{\Lambda ^4}{4 \tilde{\xi }^2} \left\{\sinh \left[\frac{\sqrt{\frac{2}{3}} \chi }{M_P} - \arcsinh\left( 2 \tilde{f_0}{}^2\right)\right]+2 \tilde{f_0}{}^2\right\}^2  \label{eq:U:app} \, .
\ee
Note that the potential~\eqref{eq:U:app} is completely independent on $\Omega(\phi)$ and can be generated by the action
\be 
S= \int d^4x\sqrt{-g}\left[\frac{M_P^2}{2} \left( {\cal R}+\tilde f_0^2 \tilde{\cal R} \right) + c \tilde {\cal R}^2 \right], \qquad  c=\tilde{\xi }^2 \left( \frac{M_P}{4 \Lambda} \right)^4 \, ,
\label{eq:S:Rtilde2} 
\ee
which has been already studied in~\cite{Salvio:2022suk}. Using the properties of the hyperbolic functions, it can be proven that eq.~\eqref{eq:U:app} is equivalent to the expression of the inflaton potential used in~\cite{Salvio:2022suk}.
Moreover, using the same properties, we can also rewrite eq. \eqref{eq:U:app} as 
\be
 U(\chi)_{\tilde {\cal R}^2} \simeq \frac{\Lambda ^4}{4 \tilde{\xi }^2} \left\{
 \sqrt{1+ 4 \tilde{f_0}{}^4} \sinh \left(\frac{\sqrt{\frac{2}{3}} \chi }{M_P}\right) 
 -  2 \tilde{f_0}{}^2 \cosh \left(\frac{\sqrt{\frac{2}{3}} \chi }{M_P}\right)
 +2 \tilde{f_0}{}^2
 \right\}^2  \label{eq:U:app:2} \, .
\ee
Therefore, taking the $\tilde f_0 \to \infty$ limit of~\eqref{eq:U:app:2}, we easily obtain the Starobinsky potential~\cite{Starobinsky:1980te}
\be
 U(\chi)_{R^2} \simeq \frac{\Lambda ^4 \tilde{f_0}{}^4 }{\tilde{\xi }^2} \left(1-e^{-\sqrt{\frac{2}{3}}\frac{ \chi }{M_P}}\right)^2  \label{eq:U:app:R2} \, .
\ee
Therefore, because of such an universal\footnote{Obviously this is not the only way to generate an Einstein frame Starobinsky potential from a MAG setup. More details about it are given in \ref{appendix:attractors}.} strong coupling limit and the analogy with the $\xi$-attractors, we decide to label the class of models defined by eqs.~\eqref{eq:attractorschoice} as \emph{$\tilde\xi$-attractors}. It is reasonable to expect that the $\tilde\xi$-attractors will show two asymptotic behaviours in two different steps, the first one approaching the predictions of the $\tilde {\cal R}^2 $ model in~\eqref{eq:S:Rtilde2} for $|\tilde\xi| \gg 1$ and then a second one approaching the results of Starobinsky inflation when also $|\tilde f_0| \gg 1$. In order to test numerically such a behaviour, from now on we study the specific choice
\be
\Omega(\phi)^2 = \left(\frac{\phi}{M_P}\right)^n \, ,\label{eq:Omega}
\ee
with $n>0$. We note that for even $n$'s  $V(\phi)$ is positive for any $\phi$ values and all the functions in~\eqref{eq:attractorschoice}  are symmetric under the transformation $\phi \to -\phi$. On the other hand, for odd $n$'s, the inflaton potential is positive only for $\phi>0$. Therefore from now on we only work in the positive quadrant for $\phi$.
Moreover, the absolute signs of $\tilde f$ and $\tilde f_0 $ are irrelevant (see eqs.~\eqref{eq:k(phi)} and~\eqref{eq:attractorschoice}), therefore from now on we choose the convention where ${\tilde f_0}>0$ while $\tilde{\xi }$ changes sign.
Inserting~\eqref{eq:Omega} into~\eqref{eq:k(phi):omega}, the kinetic function becomes
\be
   k(\phi) = 1+\frac{3 \, n^2 \, \tilde{\xi }^2 \left(\frac{\phi }{M_P}\right)^{n-2}}{2 \left[1+4 \left(\tilde{f_0}^2+\tilde{\xi } \left(\frac{\phi }{M_P}\right)^{n/2}\right)^2\right]} \, .
    \label{eq:k(phi):xi0}
\ee
 The position of the corresponding local maximum can be computed to be
\be
 \phi_\text{peak}= \left(\frac{\Delta}{\tilde{\xi }}\right)^{2/n }M_P   \qquad \Delta = \frac{\tilde{f_0}{}^2}{4} \left( n-4-\sqrt{n^2+\frac{2 (n-2)}{\tilde{f_0}{}^4}} \right)\, .
 \label{eq:peak}
\ee
and the value of the maximum of the kinetic function is
\be
 k(\phi_\text{peak}) = 1+ \tilde{\xi }^{4/n} \frac{3 n^2 \Delta ^{2-\frac{4}{n}}}{2 \left(4 \left(\tilde{f_0}{}^2+\Delta \right){}^2+1\right)}\label{eq:k:peak} \, .
\ee
Since $\Delta$ is always negative, in order to ensure that $\phi_\text{peak}$ is real and positive, $\tilde \xi$ must be negative as well. Therefore, from now on we will only consider $\tilde\xi<0$. Moreover, if $0<n<2$,  then $\tilde{f_0}$ has the lower bound $\tilde{f_0}{}^2 \geq \sqrt{2} \sqrt{\frac{2-n}{n^2}}$. Now let us see how the peak in $k(\phi)$ generates a flat region in $U(\chi)$. This happens via an inflection point, whose equation is:
\be
 U''(\chi_\text{flex})=0 \label{eq:flex:point} \, .
\ee
In terms of $\phi$, eq.~\eqref{eq:flex:point} can be rewritten as
\be 
 \frac{1}{2} \frac{k'(\phi_\text{flex})}{k(\phi_\text{flex})} = \frac{V''(\phi_\text{flex})}{V'(\phi_\text{flex})} \, , \label{eq:flex:point:2}
\ee
where we have used $\xi=0$. Now, using eqs.~\eqref{eq:attractorschoice} and~\eqref{eq:Omega}, eq.~\eqref{eq:flex:point:2} becomes
\be 
 \frac{M_P}{2} \frac{k'(\phi_\text{flex})}{k(\phi_\text{flex})} 
 = (n-1) \frac{M_P}{\phi_\text{flex}}\, . \label{eq:flex:point:3}
\ee
Then, it is easy to see that for $n=1$, eq.~\eqref{eq:flex:point:3} just becomes $k'(\phi_\text{flex})=0$ and $\phi_\text{flex} = \phi_\text{peak}$. For any other $n$, if $\phi_\text{flex} \gg M_P$, then $\phi_\text{flex} \simeq \phi_\text{peak}$. This happens when $\tilde f_0 \gg 1$ (see eq.~\eqref{eq:peak}).
Let us now study the behaviour of $U(\chi)$ in the vicinity of the pronounced peak in $k(\phi)$. From eq.~\eqref{eq:k:peak} it is easy to check that if $|\tilde\xi| \gg 1$ and/or $\tilde f_0 \gg 1$, then the value of $k(\phi)$ at the maximum behaves like
\be
k(\phi_\text{peak}) \approx \frac{3}{2} n^2 \, \tilde{\xi }^{4/n} \, \tilde{f_0}{}^{4-\frac{8}{n}} \gg 1 \, .
\ee
Moreover, using the result of~\eqref{eq:chi:omega:app} combined with eq.~\eqref{eq:Omega}, we can provide an approximated solution for the field redefinition as
\be
 \chi \simeq - \sqrt{\frac{3}{2}} M_P \left\{ \arcsinh\left[2 \left(\tilde{f_0}{}^2+\tilde{\xi } \left(\frac{\phi
   }{M_P}\right)^{n/2}\right)\right]- \arcsinh\left(2 \tilde{f_0}{}^2\right) \right\} \, ,
   \label{eq:chi:app}
\ee
which leads to the same Einstein frame potential shown in eq.~\eqref{eq:U:app}.
\begin{figure}[t]
   \centering
 \subfloat[]{\includegraphics[width=0.49\textwidth]{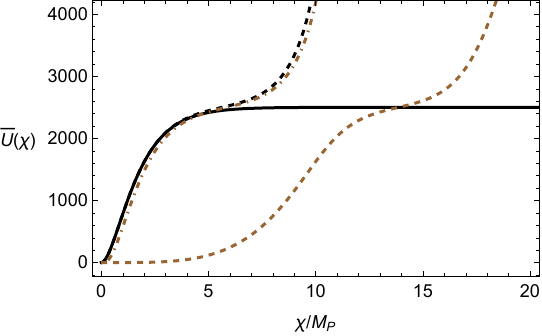}} \
  \subfloat[]{\includegraphics[width=0.5\textwidth]{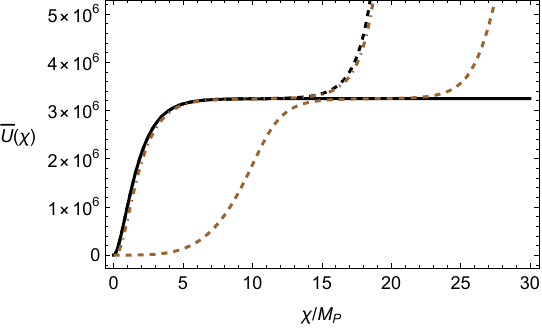}}
    \caption{$\bar U(\chi)$ vs. $\chi/M_P$ for $n=4$.  (a) $\tilde f_0=5$ with  $|\tilde\xi| \simeq 0.22$ (brown, dashed) and $|\tilde\xi| \simeq 46.8$ (brown, dot-dashed). (b) $\tilde f_0=30$ with $|\tilde\xi| \simeq 7.24$ (brown, dashed) and $|\tilde\xi| \simeq 1.66 \times 10^3$ (brown, dot-dashed). For reference the corresponding $\bar U(\chi)_{\tilde{\cal R}^2}$ (black, dashed) and $\bar U(\chi)_{R^2}$ (black, continuous).} 
\label{fig:Uplot}
\end{figure}
In order to have a better understanding of the behaviour of the inflaton potential at big $\tilde f_0$, we show in Fig.~\ref{fig:Uplot}, the plot of $\bar U(\chi)= \frac{4 \tilde \xi^2}{\Lambda^4}  U(\chi)$ (brown) for $n=4$, $\tilde f_0=5$ (a), $\tilde f_0=30$ (b) with respectively $|\tilde\xi| \simeq 0.22$ (dashed), $46.8$ (dot-dashed) and $|\tilde\xi| \simeq 7.24$ (dashed), $1.66 \times 10^3$ (dot-dashed). For reference we will also show the plots of the corresponding $\bar U(\chi)$ potentials for the $\tilde {\cal R}^2 $ model in eq.~\eqref{eq:U:app} (black, dashed) and for the Starobinsky model in eq.~\eqref{eq:U:app:R2} (black, continuous). We can see that all the brown lines in both plots exhibit a quite flat inflection point and that the corresponding concave knee if moving to smaller $\chi$'s by increasing $|\tilde\xi|$. Moreover when $|\tilde\xi|$ is large, the plateau region of $\bar U(\chi)$ is essentially indistinguishable from the one of $\bar U(\chi)_{\tilde {\cal R}^2}$. Finally, when both $|\tilde\xi|$ and $\tilde f_0$ are big, the plot of $\bar U(\chi)_{\tilde {\cal R}^2}$ overlaps the one of $\bar U(\chi)_{R^2}$ until it passes the inflection point. Then $\bar U(\chi)_{\tilde {\cal R}^2}$ turns upwards into a convex behaviour diverging to infinity while $\bar U(\chi)_{R^2}$ remains concave while approaching its horizontal asymptote.

\section{Inflationary results} \label{sec:results}

In this section we discuss the inflationary predictions of the model. As well known, in the slow-roll approximation, all the inflationary observables can be computed from the potential slow-roll parameters:
\bea
\epsilon_U  (\chi) &=& \frac{M_P^2}{2}\left(\frac{U'(\chi)}{U(\chi)}\right)^2 \, , \label{eq:epsilon}
\\
\eta_U  (\chi) &=& M_P^2 \frac{U''(\chi)}{U(\chi)} \, . \label{eq:eta}
\eea
The expansion of the Universe is estimated in number of e-folds, which is 
\be
N_e =  \frac{1}{M_P^2} \int_{\chi_{\textrm{end}}}^{\chi_N} {\rm d}\chi \, \frac{U(\chi)}{U'(\chi)} ,
\label{eq:Ne}
\ee
where the field value at the end of inflation is given by $\epsilon  (\chi_{\textrm{end}}) = 1 $, while the field value $\chi_N$ at the time a given scale left the horizon is given by the corresponding $N_e$. 
The tensor-to-scalar ratio $r$ and the scalar spectral index $n_\textrm{s}$ 
are:
\bea
r  &=& 16\epsilon_U  (\chi_N) \,  , \label{eq:r} \\
n_\textrm{s}  &=& 1+2\eta_U  (\chi_N)-6\epsilon_U  (\chi_N) \, .  \label{eq:ns} 
\eea
Finally, the amplitude of the scalar power spectrum is
\be
 A _\textrm{s} = \frac{1}{24 \pi^2 M_P^4}\frac{U(\chi_N)}{\epsilon_U  (\chi_N)}  \simeq 2.1 \times 10^{-9} \, ,
 \label{eq:As:th}
\ee
whose experimental constraint~\cite{Planck:2018jri} usually fixes the energy scale of inflation. The explicit analytical expression for the inflationary observables are too cumbersome to provide any useful information at first glance, therefore we postpone them into a separate Appendix~\ref{appendix}.
Simpler expressions can be found for the limit cases considered in Section \ref{sec:U}. The exact analytical solutions for the $\tilde {\cal R}^2$ limit configuration described by eq. \eqref{eq:U:app:2} can be found in \cite{Salvio:2022suk}. Here we just provide the corresponding approximations at the leading order in $\frac{1}{\tilde f_0}$. The results are the following
\bea
\chi_N & \approx & \sqrt{\frac{3}{2}} M_P \ln \left(\frac{4 N_e}{3}\right) \left(1+\frac{N_e^2}{27 {\tilde f_0}^4 \ln \left(\frac{4 N_e}{3}\right)}\right) \, , \label{eq:chiN:app} \\
 r &\approx & \frac{12}{N_e^2}\left(1+ \frac{4 N_e^2}{27 {\tilde f_0}^4} \right) \, , \label{eq:r:app} \\
 n_s &\approx & 1-\frac{2}{N_e}\left(1 - \frac{4 N_e^2}{27 {\tilde f_0}^4} \right)  \, , \label{eq:ns:app} \\
 A_s &\approx & \frac{\Lambda^4}{\tilde\xi^2 M_P^4} \frac{{\tilde f_0}^4 N_e^2 }{72 \pi ^2} \label{eq:As:app}  \left(1-\frac{4 N_e^2}{27 {\tilde f_0}^4}\right) \, ,
\eea
where we have also used the large $N_e$ approximation. Therefore we can expect that the $\tilde {\cal R}^2$ limit will imply larger values for both $r$ and $n_s$ with the respect to Starobinsky inflation, as shown in \cite{Salvio:2022suk}. In the limit $\tilde f_0 \to \infty$, eqs. reproduce the results of Starobinsky inflation (in the big $N_e$ approximation), as expected from eq. \eqref{eq:U:app:R2}. We stress that the phenomenological parameters in eqs. \eqref{eq:r:app}, \eqref{eq:ns:app} and \eqref{eq:As:app} exhibit all the same correction factor $\frac{4 N_e^2}{27 {\tilde f_0}^4} $. Therefore, for practical purposes, we will state that the ``Starobinsky limit is reached" when such a factor is of the order of $10^{-3}$ or less. Hence, from eq. \eqref{eq:r:app},  we expect, that the Starobinsky limit is reached when $\tilde f_0 \gtrsim 3.5 \sqrt{N_e}$.


\begin{figure}[t]
     \subfloat[]{\includegraphics[width=0.5\textwidth]{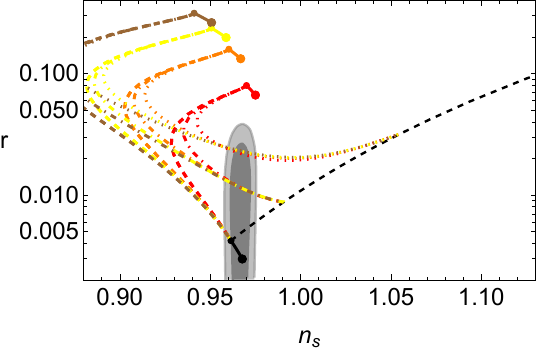}}%
\ 
    \subfloat[]{\parbox{0.5\textwidth}{\vspace*{-4.75cm}
    \includegraphics[width=0.5\textwidth]{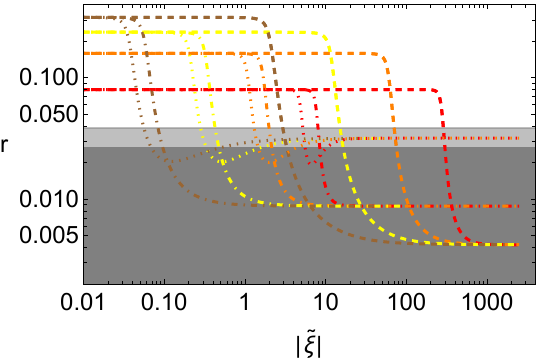}}}%

   \subfloat[]{\includegraphics[width=0.5\textwidth]{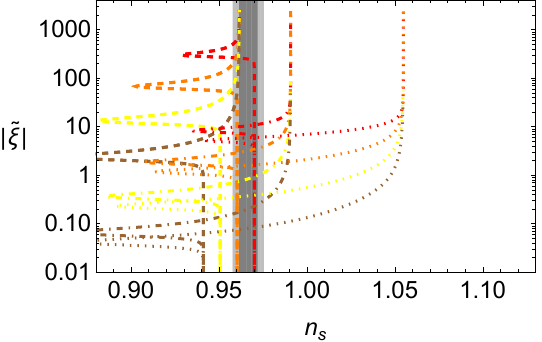}}%
    \ 
    \subfloat[]{\includegraphics[width=0.49\textwidth]{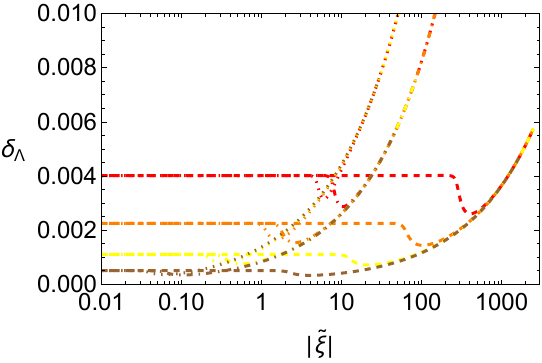}}%

    \caption{$r$ vs.~$n_\text{s}$ (a),  $r$ vs.~$|\tilde\xi|$ (b),  $|\tilde\xi|$ vs.~$n_\text{s}$ (c),  $\delta_\Lambda$ vs.~$|\tilde\xi|$ (d) for $N_e=50$ and $n=1$ (red), $2$ (orange), $3$ (yellow) and $4$ (brown) with ${\tilde f_0}=4$ (dotted), ${\tilde f_0}=5$ (dot-dashed) and ${\tilde f_0}=30$ (dashed). The gray areas represent the 1,2$\sigma$ allowed regions coming  from  the latest combination of Planck, BICEP/Keck and BAO data~\cite{BICEP:2021xfz}. For reference the predictions of ${\tilde{\cal R}}^2$ inflation (black, dashed) at $N_e=50$ and of standard $\phi^n$ inflation (continuous, same colors as before) and Starobinsky inflation (black, continuous) for $N_e \in [50,60]$.  } 
    \label{fig:rvsns:50}
\end{figure}

\begin{figure}[t]
     \subfloat[]{\includegraphics[width=0.5\textwidth]{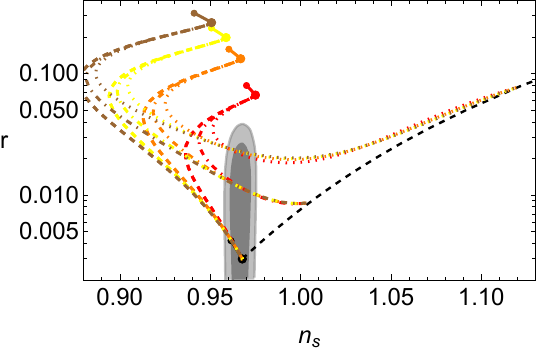}}%
    \
    \subfloat[]{\parbox{0.5\textwidth}{\vspace*{-4.75cm}
    \includegraphics[width=0.5\textwidth]{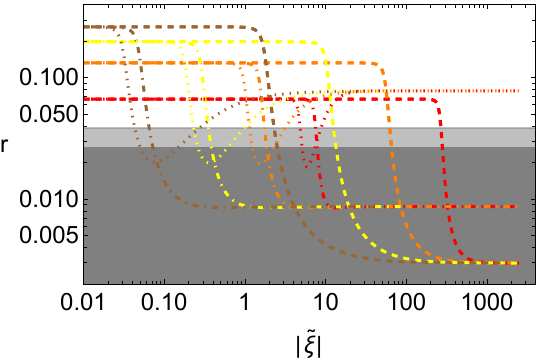}}}%
  
   \subfloat[]{\includegraphics[width=0.5\textwidth]{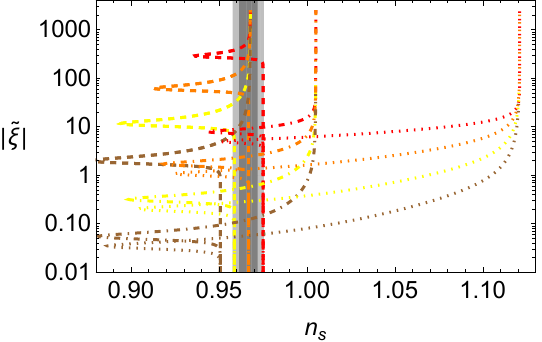}}%
    \ 
    \subfloat[]{\includegraphics[width=0.49\textwidth]{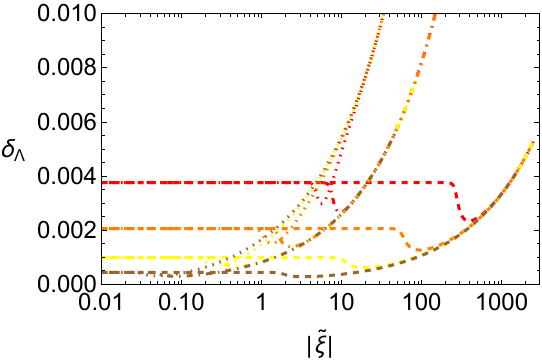}}%

    \caption{$r$ vs.~$n_\text{s}$ (a),  $r$ vs.~$|\tilde\xi|$ (b),  $|\tilde\xi|$ vs.~$n_\text{s}$ (c),  $\delta_\Lambda$ vs.~$|\tilde\xi|$ (d) for $N_e=60$ and $n=1$ (red), $2$ (orange), $3$ (yellow) and $4$ (brown) with ${\tilde f_0}=4$ (dotted), ${\tilde f_0}=5$ (dot-dashed) and ${\tilde f_0}=30$ (dashed). The gray areas represent the 1,2$\sigma$ allowed regions coming  from  the latest combination of Planck, BICEP/Keck and BAO data~\cite{BICEP:2021xfz}. For reference the predictions of ${\tilde{\cal R}}^2$ inflation (black, dashed) at $N_e=60$ and of standard $\phi^n$ inflation (continuous, same colors as before) and Starobinsky inflation (black, continuous) for $N_e \in [50,60]$.  } 
    \label{fig:rvsns:60}
\end{figure}

Let us now investigate the model away from the attractor configurations.
The corresponding numerical results are instead illustrated in Fig.~\ref{fig:rvsns:50} for $N_e=50$ and Fig.~\ref{fig:rvsns:60} for $N_e=60$, where we show  $r$ vs.~$n_\text{s}$ (a),  $r$ vs.~$|\tilde\xi|$ (b),  $|\tilde\xi|$ vs.~$n_\text{s}$ (c),  $\delta_\Lambda$ vs.~$|\tilde\xi|$ (d) with $\delta_\Lambda = \frac{\Lambda}{M_P}$ counting the prefactor of the inflaton potential~\eqref{eq:attractorschoice}  in Planck units. We considered the following values for $n$: $1$ (red), $2$ (orange), $3$ (yellow) and $4$ (brown).  In both Figs.~\ref{fig:rvsns:50} and~\ref{fig:rvsns:60} we used ${\tilde f_0}=4$ (dotted), ${\tilde f_0}=5$ (dot-dashed) and ${\tilde f_0}=30$ (dashed).  For reference we also added the 1,2$\sigma$ allowed regions coming  from  the latest combination of Planck, BICEP/Keck and BAO data~\cite{BICEP:2021xfz} (gray areas) and the predictions of ${\tilde{\cal R}}^2$ inflation (black, dashed) at $N_e=50,60$ (according to the figure) and of standard $\phi^n$ inflation (continuous, same colors as before) and Starobinsky inflation (black, continuous) for $N_e \in [50,60]$.  

First all we notice that, as usual, moving from $N_e=50$ (Fig.~\ref{fig:rvsns:50}) to $N_e=60$ (Fig.~\ref{fig:rvsns:60}), the predictions move towards lower (higher) $r$ ($n_s$) but the generic behaviour remains unaffected. However, the lower ${\tilde f_0}$, the higher is the increase in $n_s$ at a given $\tilde\xi$ (cf. Figs.~\ref{fig:rvsns:50}(c) and~\ref{fig:rvsns:60}(c)). 
As expected from the analytical study of the inflaton potential and the limits in eqs.~\eqref{eq:U:app} and~\eqref{eq:U:app:R2}, we see that for $|\tilde\xi| \gg 1$ we predictions are aligned with the ones of the $\tilde {\cal R}^2$ model. When also $\tilde f_0 \gg 1$ (specifically $\tilde f_0=30$ from our list of benchmark points), the results reach the Starobinsky limit\footnote{This is in agreement with the results of~\cite{Salvio:2022suk}, where, using our notation, the author just stopped the numerical analysis at ${\tilde f_0} = 10\sqrt3 \simeq 17.3$.  } as expected from eqs. \eqref{eq:r:app}  and \eqref{eq:ns:app}.
 Even though it is possible to get predictions in agreement with the latest constraints~\cite{BICEP:2021xfz}, most of the predictions actually fall out of the 2$\sigma$ allowed region. On the other hand, it is possible to reach the 1$\sigma$ region without using very large $\tilde f_0$ but just $\tilde f_0 = 4$. 

From Figs.~\ref{fig:rvsns:50}(b) and~\ref{fig:rvsns:60}(b), we see that $r$ is insensitive to the  value of $\tilde\xi$ when $|\tilde\xi|<1$, for all the considered values of $\tilde f_0$. By increasing $|\tilde\xi|$, $r$ decreases until it reaches its asymptotic value corresponding to $\tilde {\cal R}^2$ inflation, for all the studied values of $\tilde f_0$ but $\tilde f_0=4$, where instead, it first reaches a minimum and then increases towards the aforementioned limit. Finally we notice that the generic shape of the results is not affected by changing $n$, but the asymptotic configuration is reached at larger $|\tilde\xi|$ values with $n$ increasing.

From Figs.~\ref{fig:rvsns:50}(c) and~\ref{fig:rvsns:60}(c), we see that behaviour of $n_s$ is the same for all the considered values of $\tilde f_0$. First, it is insensitive to the  value of $\tilde\xi$ when $|\tilde\xi|<1$. By increasing $|\tilde\xi|$, $n_s$ decreases until it reaches a minimum and then it increases reaching its asymptotic value corresponding to $\tilde {\cal R}^2$ inflation. As before, the generic shape of the results is not affected by changing $n$, but the asymptotic configuration is reached at larger $|\tilde\xi|$ values with $n$ increasing.

Finally, from Figs.~\ref{fig:rvsns:50}(d) and~\ref{fig:rvsns:60}(d), we see that behaviour of $\delta_\Lambda$ is the same for all the considered values of $\tilde f_0$. First, it is insensitive to the  value of $\tilde\xi$ when $|\tilde\xi|<1$. By increasing $|\tilde\xi|$, $\delta_\Lambda$ decreases until it reaches a minimum and then it increases reaching its asymptotic configuration corresponding to $\tilde {\cal R}^2$ inflation. Again, the generic shape of the results is not affected by changing $n$, but the asymptotic configuration is reached at larger $|\tilde\xi|$ values with $n$ increasing.

Before concluding, we briefly compare the predictions of the $\tilde \xi$-attractors with the ones of the $\xi$-attractors \cite{Kallosh:2013tua,Galante:2014ifa}, since both models exhibit Starobinsky inflation as an attractor solution. Certainly the $\xi$-attractors setup is simpler because it requires only one parameter, $\xi$, in addition to the normalization of scalar potential $\Lambda$, while the $\tilde \xi$-attractors require two, $\tilde f_0$ and $\tilde \xi$. Moreover the attractor behaviour of \cite{Kallosh:2013tua,Galante:2014ifa} is stronger because, for $n \geq 2$, it occurs already when $\xi < 1$, while the $\tilde \xi$-attractors are insensitive to the $\tilde \xi$ value for $|\tilde \xi| < 1$. Nevertheless, $\tilde \xi$-attractors provide an interesting scenario since, at a given $\Omega$, cover a different region of the $r$ vs. $n_s$ plane when away from the limit configurations.

To conclude we note that our results agree with the ones of~\cite{Langvik:2020nrs} in the corner of the parameters space where the two models are comparable. However, no direct comparison can be done with~\cite{Shaposhnikov:2020gts} because they do not study the case of a negative $\tilde\xi$.

\subsection{On the validity of the attractor solutions}
\label{subsec:validity}

\begin{figure}[t]
   \centering
    \subfloat[]{\includegraphics[width=0.48\textwidth]{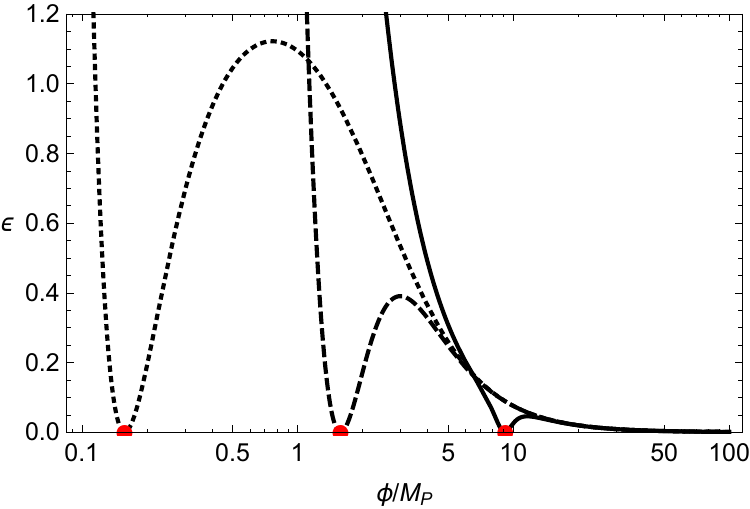}} \
  \subfloat[]{\includegraphics[width=0.5\textwidth]{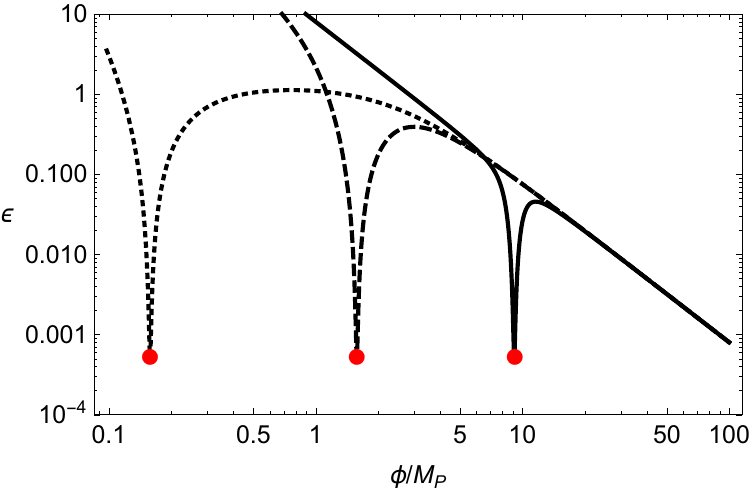}}
         \caption{(a) $\epsilon(\phi)$ for $n=4$, $\tilde f_0=5 $ and $|\tilde\xi|=0.3$ (continuous), $|\tilde\xi|=10$ (dashed) and $|\tilde\xi|=1000$ (dotted). The red bullets represents the corresponding $(\phi_N,\epsilon(\phi_N))$ coordinates. (b) the same but with a logarithmic scale for $\epsilon$.} 
    \label{fig:eps}
\end{figure}

In this subsection we investigate in more details the validity of our findings. Without loss of generality we focus initially on the $n=4$ case, and later we will extend the analysis to the remaining $n$ values.
First of all, we provide in Fig. \ref{fig:eps}(a) a plot of the first-slow roll parameter $\epsilon$ given in eq. \eqref{eq:epsilon} (see also eq.~\eqref{eq:epsilon:x:phi}) as a function of $\phi$ for $n=4$, $\tilde f_0=5 $ and three different $\tilde\xi$ values that they all reproduce the asymptotic limit of $r$ at $N_e=60$ (see Fig. \ref{fig:rvsns:60}): $|\tilde\xi|=0.3$ (continuous), $|\tilde\xi|=10$ (dashed) and $|\tilde\xi|=1000$ (dotted). The red bullets represents the corresponding $(\phi_N,\epsilon(\phi_N))$ coordinates. Fig. \ref{fig:eps}(b) is the same as Fig. \ref{fig:eps}(a) but with a logarithmic scale for the $\epsilon$ values. First of all we notice that when $\phi \gtrsim 20 M_P$, all the three lines converge to the same behaviour. This happens because in the $\phi \to +\infty$ limit, the kinetic function behaves like
\be
   k(\phi) \simeq  1+\frac{3 M_P^2}{2}  \, \left[\frac{\Omega(\phi)'}{  \Omega(\phi) }\right]^2 \,  
   = 1+\frac{3 \, n^2 \, }{8} \left(\frac{M_P}{\phi }\right)^{2} =  1+6 \left(\frac{M_P}{\phi }\right)^{2}\,  ,
    \label{eq:k(phi):omega:app:naive}
\ee
where we have used eq. \eqref{eq:Omega} and $n=4$. Note that both the $\tilde f_0$ and $\tilde\xi$ contributions are absent. Moreover for $\phi \gg M_P$, $k(\phi) \approx 1$ and the setup behaves like standard monomial inflation. However, this is not our region of interest. We see from Fig. \ref{fig:eps} that $\epsilon$ always develops a local maximum in $\phimax$ and  a local minimum in $\phimin$ with $\phi_N \simeq \phimin$. On one hand $\epsilon(\phi_N)$ is always the same as expected, since we chose the points in the asymptotic limit of $r$. On the other hand, with $|\tilde\xi|$ increasing, $\phimax$ decreases and $\epsilon(\phimax)$ increases.  In particular for $|\tilde\xi|=1000$, we have $\epsilon(\phimax)>1$. This implies having two different regions available for inflation: one before and one after $\phimax$. This puts in jeopardy the validity of the predictions\footnote{Such a scenario might still be interesting in the context of primordial black holes production (e.g. \cite{Dimopoulos:2019wew,Karam:2023haj} and refs. therein). However this exceeds our present purpose and we postpone such a study to a future work.}, because either we have very fine-tuned initial conditions where the initial $\phi$ value is smaller than $\phimax$ or the result is not acceptable because it does not take into take into account the first inflationary stage when $\phi$ starts to roll down from values larger than $\phimax$. Therefore, the predictions for $|\tilde\xi|=1000$ should be somehow rejected. 

An analogous situation repeats for any $n$ and $\tilde f_0$ values when $|\tilde\xi|$ is big enough. Therefore, we could set a naive threshold value on $|\tilde\xi|$, by requiring that $\epsilon(\phimax)<1$. In order to do so, first of all we need to solve the equation for the stationary points of $\epsilon$, which is quite cumbersome. A more readable result can be obtained using the approximation $\tilde f_0^4 \gg 1$, which is satisfied by all our benchmark points, obtaining
\be
\left(\frac{\tilde{f_0}{}^2}{\tilde{\xi }}+ x_{\phi }^{n/2}\right) \left[16 x_{\phi }^{6/n} +\frac{3 \tilde{f_0}{}^2 }{\tilde{\xi }}\left(n^3+16 x_{\phi }^{\frac{8}{n^2}}\right) x_{\phi }^{\frac{4 (n-2)}{n^2}}+\frac{48 \tilde{f_0}{}^4 x_{\phi }^{2/n}}{\tilde{\xi }^2}+\frac{16
   \tilde{f_0}{}^6}{\tilde{\xi }^3}\right] =0 \, \label{eq:eps:stat}
\ee
 with $x_\phi = \frac{\phi}{M_P}$. The term in the round brackets gives the minimum solution
 \be 
 \phimin=\left(\frac{\tilde{f_0}^2}{|\tilde\xi|}\right)^{2/n} M_P \, , \label{eq:phi:min}
 \ee
  which coincides with $\phi_\text{peak}$ in \eqref{eq:peak} in the $\tilde f_0^4 \gg 1$ approximation, confirming that $\phi_N \simeq \phi_\text{peak}$. The term in the square brackets provides the equation for $\phimax$, for which we cannot provide a general\footnote{It is actually possible to find exact solutions for $n=2,4$, but the exact result would not add much to the discussion, therefore we omit it for the sake of simplicity of the analysis.} analytical solution for an arbitrary $n$. However, we can perform some approximations that will help us to reach a rough upper bound on $|\tilde\xi|$. We tested numerically that if $\epsilon(\phimax) \gtrsim 1$, then $\frac{\tilde{f_0}{}^2}{|\tilde \xi }| \ll 1$ and $x_{\phimax} <1$. Therefore the maximum equation can be written at the leading order in $\frac{\tilde{f_0}{}^2}{\tilde \xi } $ and $x_{\phi}$ as
  \be
16 x_{\phi }^{6/n} +\frac{3 \tilde{f_0}{}^2 }{\tilde{\xi }} n^3 x_{\phi }^{\frac{4 (n-2)}{n^2}} =0 \, , \label{eq:eps:max}
\ee
which can be solved, giving
  \be
\phimax \simeq \left(\frac{3}{16}\frac{n^3 \tilde{f_0}{}^2}{|\tilde{\xi }|}\right)^{\frac{2}{n+4}}  M_P \, . \label{eq:phi:max}
\ee
Using similar approximations, we can also solve $\epsilon(\phimax)<1$, obtaining the rough upper bound
\be
 | \tilde\xi | < \tilde\xi_\text{max} = 2 \left[\left(\frac{32}{3}\right)^n n^{-3 n} (n+4)^{n+4}\right] ^{1/4} \tilde{f_0}{}^2 \simeq \left\{
 \begin{array}{lcr}
 27.02 \tilde{f_0}{}^2 \, , & & n=1 \\
 33.94 \tilde{f_0}{}^2 \, ,  & & n=2 \\
 30.02 \tilde{f_0}{}^2 \, ,  & & n=3 \\
 21.33 \tilde{f_0}{}^2 \, ,  & & n=4  
 \end{array}
 \right.
 \, , \label{eq:xitilde:max}
\ee
which is roughly in agreement with the assumption  $\frac{\tilde{f_0}{}^2}{| \tilde \xi | } \ll 1$. Going back to the $n=4$, $\tilde{f_0}=5$ example, we see that the corresponding upper bound is $|\tilde\xi| \lesssim 533.25$, in agreement with what depicted in Fig. \ref{fig:eps}. Moreover, we note that in all the configurations considered, both the $\tilde {\cal R}^2$ and the Starobinsky limit are reached before the insurgence of the upper bound in $| \tilde\xi |$. Therefore the attractor solution is confirmed, but only within the validity of \eqref{eq:xitilde:max}.

To conclude, we comment on the eventual impact of the  non-minimal function $f(\phi)$ given in eq. \eqref{eq:attractorschoice}. As we mentioned before, such a contribution should imply a loss of attractor behaviour. On the other hand, $\xi \neq 0$ becomes relevant at big $\phi$ values, while our relevant inflationary region is at small $\phi$'s (see eqs. \eqref{eq:peak} and \eqref{eq:phi:min}). Therefore, the attractor behaviour of our model will be not spoiled if the condition $\xi \Omega(\phi_N) \ll 1$ is satisfied. We can estimate such an upper bound as
\be
 \xi \ll \left(\frac{M_P}{\phi_N} \right)^{n/2} \simeq \frac{\tilde \xi}{\tilde{f_0}{}^2} \lesssim  2 \left[\left(\frac{32}{3}\right)^n n^{-3 n} (n+4)^{n+4}\right] ^{1/4}  \simeq \left\{
 \begin{array}{lcr}
 27.02  \, , & & n=1 \\
 33.94  \, ,  & & n=2 \\
 30.02  \, ,  & & n=3 \\
 21.33  \, ,  & & n=4  
 \end{array}
 \right. \, , \label{eq:xi:max}
\ee
where we have used eq. \eqref{eq:Omega}, the approximation $\phi_N \simeq \phi_\text{min}$, eq. \eqref{eq:phi:min}, $\tilde \xi \lesssim \tilde \xi_\text{max}$ and eq. \eqref{eq:xitilde:max}. Given the results of eq. \eqref{eq:xi:max}, we can expect that the attractor behaviour described before for $1 \leq n \leq 4$ will not be compromised by the presence of a $f(\phi) \neq 1$, as long as $\xi \lesssim 0.1$. More details on attractors solutions in presence of both $f(\phi)$ and $\tilde f(\phi)$ are also given in Appendix \ref{appendix:attractors}.

\section{Conclusions} \label{sec:conclusions}

We studied a new class of inflationary attractors in metric-affine gravity. Such class exhibits a non-minimal coupling function, $\Omega(\phi)$, with the Holst invariant $\tilde {\cal R}$ and an inflaton potential proportional to $\Omega(\phi)^2$. Because of the analogy  with the renown $\xi$-attractors, we decided to label this new class as $\tilde\xi$-attractors. The attractor behaviour of the class takes place with two combined strong coupling limits. The first limit is the obvious $\tilde\xi \gg 1$, which makes the theory equivalent to a $\tilde {\cal R}^2$ model. Then, the second limit considers a very small Barbero-Immirzi parameter (i.e. $\tilde f_0 \gg 1$), driving the inflationary predictions of the model into the ones of Starobinsky inflation. We also performed a detailed numerical study for $\Omega (\phi)^2 = \left( \phi/M_P \right)^n$ with $n=1,2,3,4$. The two-steps attractor behaviour has been confirmed for all the considered values of $n$. The Starobinsky limit is reached when $\tilde f_0\gtrsim 3.5 \sqrt{N_e}$ ($\tilde f_0=30$ from the list of our benchmark points). On the other hand, compatibility with experimental data~\citep{BICEP:2021xfz} at 1$\sigma$ level, is already possible for $\tilde f_0=4$ and $\tilde\xi \simeq 6,1.5,0.3,0.08,$ respectively for $n=1,2,3,4$, but far away from the Starobinsky solution. 
The forthcoming experiments with a precision of $\delta r \sim 10^{-3}$, such as Simons Observatory~\cite{SimonsObservatory:2018koc}, CMB-S4~\cite{Abazajian:2019eic} and LITEBIRD~\cite{LiteBIRD:2020khw}, will be capable to confirm or rule out our scenario, in particular for the cases away from the Starobinsky limit.

\acknowledgments
The author would like to thank I. D. Gialamas and A. Salvio for useful discussions. This work was supported by the Estonian Research Council grants PRG1055,  RVTT3, RVTT7 and the CoE program TK202 ``Foundations of the Universe". This article is based upon work from the COST Actions CosmoVerse CA21136 and BridgeQG CA23130 supported by COST (European Cooperation in Science and Technology).

\appendix

\section{General discussion on attractors} \label{appendix:attractors}
In this Appendix we provide a more general discussion about attractor solutions in MAG. First of all it is worth noticing that the MAG action (with both non-metricity and torsion present) \eqref{eq:Sstart} can be rewritten into the Palatini action (with only non-metricity present)
\be 
S_{\rm J}= \int d^4x\sqrt{-g}\left[\frac{M_P^2}{2}  \bar f(\phi){\cal R} -b(\phi) \frac{\partial_\mu \phi \, \partial^\mu \phi}{2} - \bar V(\phi) \right], 
\label{eq:Sstart:Palatini} 
\ee
where 
\be
 b(\phi) =    1 +\frac{6 M_P^2 \left[f(\phi)' {\tilde f(\phi)}-f(\phi){\tilde f(\phi)}'\right]^2}{f(\phi) \left[ f(\phi)^2+4{\tilde f(\phi)}^2 \right]} \, , \label{eq:b:phi} 
\ee
while $\bar f(\phi)=f(\phi)$ and $\bar V(\phi)=V(\phi)$. 
Therefore, using $\bar f(\phi)$, $b(\phi)$ and $\bar V(\phi)$ instead of $f(\phi)$, $\tilde f(\phi)$ and $V(\phi) $ as a fundamental functions, one can study the phenomenology of action  \eqref{eq:Sstart} by using the Palatini scalar-tensor theory in action \eqref{eq:Sstart:Palatini}. It is convenient then, to also provide the equations for the Einstein frame action in the notation of \eqref{eq:Sstart:Palatini}:
\be 
S_{\rm E} =\int d^4x\sqrt{-g}\left[\frac{\MP^2}{2} R -\frac{1}{2}\partial_\mu \chi \, \partial^\mu \chi - U(\chi) \right], \label{eq:S:E:other}
\ee 
where
\be
 U(\chi)=\frac{  \bar V(\phi(\chi))}{\bar f^2(\phi(\chi))} \, , \label{eq:U:other}
\ee
and the canonical normalized scalar is defined by solving 
\be
  \left( \frac{d\chi}{d\phi} \right)^2= \frac{b(\phi) }{\bar f(\phi)} \, . \label{eq:chi:other}
\ee
 Hence, it is immediate to see (e.g. \cite{Jarv:2020qqm} and refs. therein) that the configuration introduced in Section \ref{sec:model} is not the only one that produces the results of Section \ref{sec:results}. For instance, given an arbitrary continuous and strictly positive function  $\bar f(\phi)$ and 
\bea
 b(\phi) &=& \bar f(\phi) \, k(\phi) \, , \label{eq:b:phi:example} \\
 \bar V(\phi) &=& \bar f(\phi)^2 \, V(\phi) \, , \label{eq:Vbar:phi:example}
\eea
with $k(\phi)$ given in eq. \eqref{eq:k(phi):omega} and $V(\phi)$ given in \eqref{eq:attractorschoice}, it is easy to prove that the Einstein frame potential \eqref{eq:U:other} and scalar normalization \eqref{eq:chi:other} are respectively exactly the same as \eqref{eq:U} and \eqref{eq:k(phi)}.
Clearly, among the infinite possibilities provided by eqs. \eqref{eq:b:phi:example} and \eqref{eq:Vbar:phi:example},  $\bar f=1$ is the simplest the choice.

\section{Equations for inflationary parameters} \label{appendix}
In this Appendix we give the analytical expressions of the slow-roll parameters and the corresponding inflationary observables. Since the field redefinition~\eqref{eq:k(phi)} cannot be solved analytically, as customary, we apply the chain rule of derivatives and express the parameters in function of $\phi$. Therefore, using eqs.~\eqref{eq:U},~\eqref{eq:k(phi)},~\eqref{eq:epsilon} and~\eqref{eq:eta}, we get
\bea
 \epsilon_U(x_\phi) &=& \frac{n^2}{2 x_\phi ^2} \frac{1}{1+\frac{3 n^2 \tilde{\xi }^2 x_\phi^{n-2}}{2 \left[1+4
   \left(\tilde{f_0}{}^2+\tilde{\xi } x_\phi^{n/2}\right)^2\right]}}  \,  \label{eq:epsilon:x:phi} ,\\   
&&\quad\nn\\
   \eta_U(x_\phi) &=& \frac{1}{\left(16 \tilde{f_0}{}^2 \tilde{\xi } x_{\phi }^{\frac{n}{2}+2}+2 \left(4 \tilde{f_0}{}^4+1\right) x_{\phi }^2+\tilde{\xi }^2
   \left(3 n^2+8 x_{\phi }^2\right) x_{\phi }^n\right)^2} \times \nn\\
   && \times \bigg[
   12 n^4 \tilde{f_0}{}^4 \tilde{\xi }^2 x_{\phi }^n+36 n^4 \tilde{f_0}{}^2 \tilde{\xi }^3 x_{\phi }^{3 n/2}
   +3 n^4 \tilde{\xi }^2 x_{\phi }^n+24 n^4 \tilde{\xi }^4 x_{\phi }^{2 n} +   \nn\\
&& \qquad     + 4 (n-1) n x_{\phi}^2 \left(8 \tilde{f_0}{}^2 \tilde{\xi } x_{\phi }^{n/2}+4 \tilde{f_0}{}^4+4 \tilde{\xi }^2 x_{\phi }^n+1\right)^2 \bigg] \, ,   
\eea
where we have defined $x_\phi = \phi/M_P$. Hence,  the number of $e$-folds is computed using~\eqref{eq:Ne} as
\be
 N_e = \left[\frac{x_{\phi }^2}{2 n}+\frac{3}{8} \ln \left(1+4 \left(\tilde{f_0}{}^2+\tilde{\xi } x_{\phi }^{n/2}\right)^2\right)-\frac{3}{2} \tilde{f_0}{}^2 \arctan \left(2
   \left(\tilde{f_0}{}^2+\tilde{\xi } x_{\phi }^{n/2}\right)\right)\right]_{x_{\phi_\text{end}}}^{x_{\phi_N}}   \, ,
\ee 
while the tensor-to-scalar ratio~\eqref{eq:r} and the scalar spectral index~\eqref{eq:ns} become respectively 
\bea
 r &=& \frac{8 n^2}{x_{\phi_N} ^2} \frac{1}{1+\frac{3 n^2 \tilde{\xi }^2 x_{\phi_N}^{n-2}}{2 \left[1+4
   \left(\tilde{f_0}{}^2+\tilde{\xi } x_{\phi_N}^{n/2}\right)^2\right]}} \, , \\
&&\quad\nn\\
 n_s &=&1-\frac{n (n+2)}{x_{\phi_N}^2}+ \frac{3 n^4 \tilde{\xi }^2 x_{\phi_N}^{n-2} \left(\tilde{\xi }^2 x_{\phi_N}^n \left(3 (n-4) n+8 x_{\phi_N}^2\right)-2 \left(4
   \tilde{f_0}{}^4+1\right) x_{\phi_N}^2\right)}{2 \left(16 \tilde{f_0}{}^2 \tilde{\xi } x_{\phi_N}^{\frac{n}{2}+2}+2 \left(4
   \tilde{f_0}{}^4+1\right) x_{\phi_N}^2+\tilde{\xi }^2 \left(3 n^2+8 x_{\phi_N}^2\right) x_{\phi_N}^n\right){}^2}\nn\\
   &&+\frac{3 n^3 (n+8) \tilde{\xi }^2
   x_{\phi_N}^n}{4 x_{\phi_N}^4 \left(4 \left(\tilde{f_0}{}^2+\tilde{\xi } x_{\phi_N}^{n/2}\right){}^2+1\right)+6 n^2 \tilde{\xi }^2 x_{\phi_N}^{n+2}} \, .
\eea
To conclude, the amplitude of scalar perturbations~\eqref{eq:As:th} is given by
\be
 A_s = \frac{x_{\phi_N}^n}{24 \pi ^2} \frac{\Lambda ^4}{M_P^4}  \left[\frac{2 x_{\phi_N}^2}{n^2}+\frac{3 \tilde{\xi }^2 x_{\phi_N}^n}{1+4 \left(\tilde{f_0}{}^2+\tilde{\xi } x_{\phi_N}^{n/2}\right)^2}\right] \, .
\ee

\bibliographystyle{JHEP}
\bibliography{references}

\end{document}